\documentclass[aps,footinbib,prl,amsmath,amssymb,twocolumn,superscriptaddress,10pt]{revtex4-1}

\usepackage[utf8]{inputenc}
\usepackage[T1]{fontenc}
\usepackage[german,english]{babel}

\usepackage{amsmath}
\usepackage{mathtools}
\usepackage{amssymb}
\usepackage{amsthm}
\usepackage{microtype}
\usepackage{quoting} 
\quotingsetup{font=small}

\usepackage[pdftex]{graphicx}
\usepackage{fancyhdr}
\usepackage{bm}
\usepackage{comment}
\usepackage{booktabs}
\usepackage{eurosym} 
\usepackage{braket}
\usepackage{emptypage}
\usepackage{epstopdf}
\usepackage{verbatim}
\usepackage{enumitem}

\newcommand{\beq}{\begin{equation}}
\newcommand{\eeq}{\end{equation}}

\DeclarePairedDelimiter{\abs}{\lvert}{\rvert}

\renewcommand{\sim}{\thicksim}

{\left\lbrace\begin{array}{@{}l@{}}}%
{\end{array}\right.}

\usepackage[bookmarks=true,colorlinks,linkcolor=blue,urlcolor=blue,citecolor=blue]{hyperref} 

\begin{document}

\title{Quasi-localized excitations induced by long-range interactions \\in translationally-invariant quantum spin chains} 

\author{Alessio Lerose} 
\affiliation{SISSA --- International School for Advanced Studies, via Bonomea 265, I-34136 Trieste, Italy}
\affiliation{INFN --- Istituto Nazionale di Fisica Nucleare, Sezione di Trieste, I-34136 Trieste, Italy}

\author{Bojan \v{Z}unkovi\v{c}} 
\affiliation{Department of Physics, Faculty of Mathematics and Physics, University of Ljubljana, Jadranska 19, 1000 Ljubljana, Slovenia}

\author{Alessandro Silva} 
\affiliation{SISSA --- International School for Advanced Studies, via Bonomea 265, I-34136 Trieste, Italy}

\author{Andrea Gambassi} 
\affiliation{SISSA --- International School for Advanced Studies, via Bonomea 265, I-34136 Trieste, Italy}
\affiliation{INFN --- Istituto Nazionale di Fisica Nucleare, Sezione di Trieste, I-34136 Trieste, Italy}

\date{\today} 

\begin{abstract}
We show that long-range ferromagnetic interactions in quantum spin chains can induce spatial quasi-localization of topological magnetic defects, i.e., domain-walls, even in the absence of quenched disorder. Utilizing matrix-product-states numerical techniques, we study the non-equilibrium evolution of initial states with one or more domain-walls under the effect of a transverse field in variable-range quantum Ising chains. 
Upon increasing the range of these interactions, we demonstrate the occurrence of a sharp transition characterized by the suppression of spatial diffusion of the excitations during the accessible time scale: the excess energy density remains localized around the initial  position of the domain-walls. 
This quasi-localization is accurately reproduced by an effective semiclassical model, which elucidates the crucial role that long-range interactions play in this phenomenon.
The predictions of this work can be tested in current experiments with trapped ions.
\end{abstract}

\maketitle

Understanding transport, spreading of information and propagation of perturbations is an important research direction in the context of quantum many-body physics. In conventional non-equilibrium setups involving local gradients of conserved quantities, their hydrodynamical evolution is typically described by a diffusion law~\cite{mesoscopic1,mesoscopic2}, while quantum correlations and entanglement spread ballistically at a characteristic speed~\cite{CC_QuasiparticlePicture,KimHuse}. 
However, the presence of strong quenched disorder provides a robust mechanism, known as many-body localization~\cite{AndersonLocalization,BAA,HuseNandkishore,AltmanMBL}, for the complete suppression of transport and the dramatic slow-down of quantum information spreading~\cite{ProsenMBL,BardarsonPollmanMoore,VasseurMooreReview}. 

A long-standing and debated problem in this field is the possible occurrence of localization phenomena in systems without disorder and their characterization.
A variety of mechanisms have been proposed in this context, including configurational (thermal) disorder in the presence of strong interactions~\cite{DeRoeck1,Schiulaz1,Cirac:QuasiMBLWithoutDisorder,CarleoGlassy}, sufficiently complex multi-body~ \cite{AbaninProsen:SlowDynamics} or frustrated~\cite{QuasiMBLFrystration} couplings, strong electric fields \cite{Refael:MBLBlochOscillations,Pollmann:MBLBlochOscillations} and quantum confinement~\cite{KCTC,MazzaTransport,GorshkovConfinement}. In such cases, it has been argued that a non-equilibrium transient can arise, during which the features of the dynamical evolution are reminiscent of many-body localization. 

\begin{figure}[t]
\includegraphics[width=0.46\textwidth]{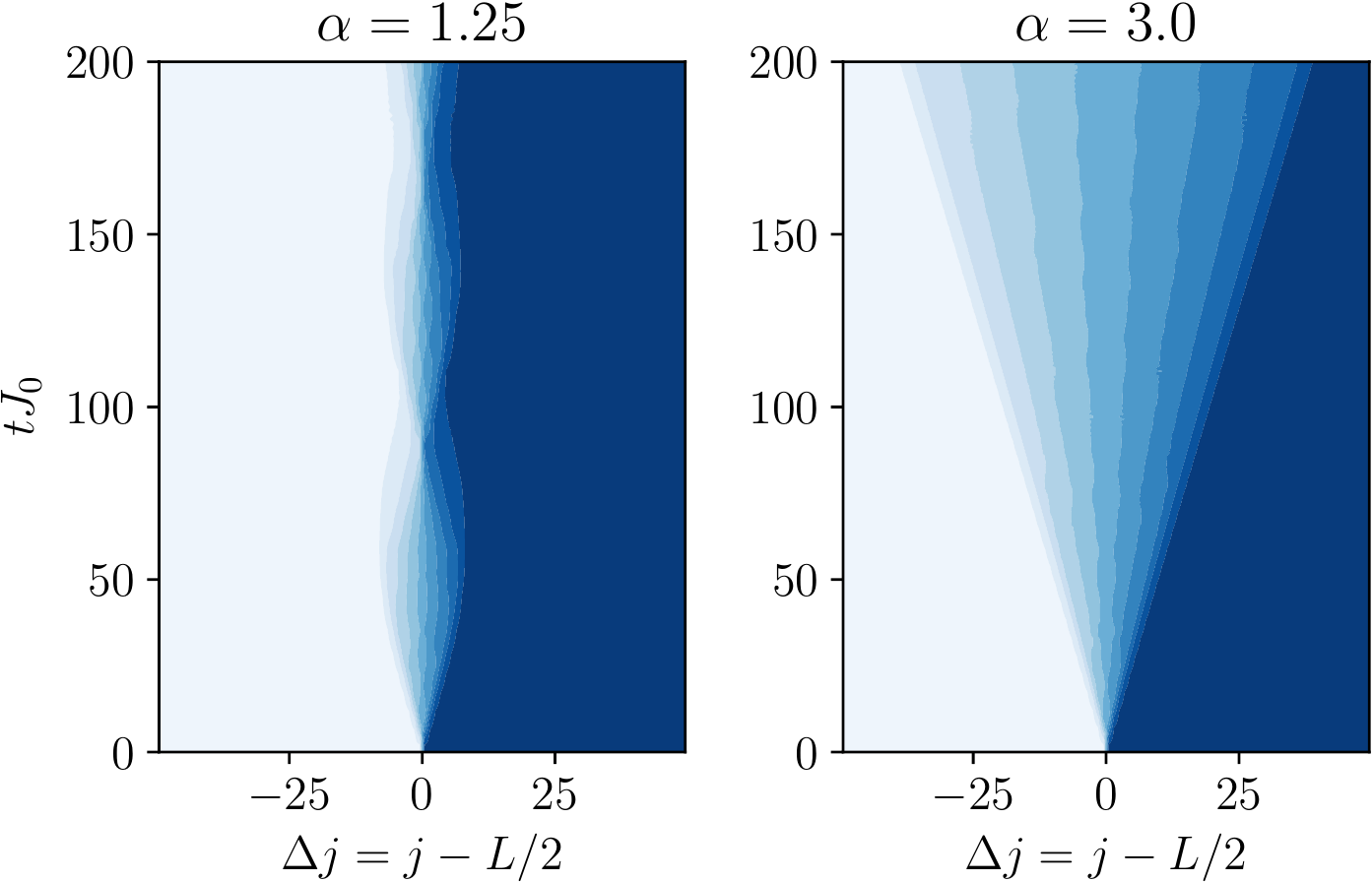} 
\caption{
Non-equilibrium evolution of the longitudinal magnetization $\langle\sigma^x_j(t)\rangle$ [cf. Eq.~\eqref{eq:H}] in an open ferromagnetic quantum Ising chain of $L=100$ spins with interactions between spins at site $i$ and $j$ given by $J_{i,j} \propto \abs{i-j}^{-\alpha}$ and $\alpha=1.25$ (left) and $3$ (right) after a global quench of the transverse field $h$ starting from a state with a single domain-wall at the center $j=L/2$ of the chain. Shown data range from $-1$ (darkest) to  $+1$ (lightest). Similar qualitative behaviors are found for $\alpha>2$ and $\alpha<2$, respectively. Data are obtained via MPS-TDVP simulations converged with bond dimension $D=64$ for the Hamiltonian \eqref{eq:H} with the quench $h=0 \to h=0.1 \, J_0$.
}
\label{fig:stmagn}
\end{figure}

In this work, we 
show that long-lived localized excitations (hereafter referred to as \emph{quasi-localized} excitations) can be observed in non-disordered quantum systems 
due to long-range interactions, relevant to experimental platforms such as trapped ions~\cite{BrittonTrappedIons,RichermeTrappedIons,JurcevicTrappedIons,ZhangDPT}, 
polar molecules~\cite{PolarMoleculesExp1,PolarMoleculesExp2},  and  magnetic atoms~\cite{MagneticAtoms,DipolarGasExp}.
Specifically, 
we study the non-equilibrium evolution of a variable-range ferromagnetic quantum Ising chain prepared in
initial states with one or more topological magnetic defects, i.e., domain-walls separating regions of uniformly magnetized spins. As illustrated in Fig.~\ref{fig:stmagn}, while in short-range systems these domain-walls undergo unbounded spatial spreading (right panel), our extensive numerical computations based on the time-dependent variational principle on matrix-product-states 
demonstrate the absence of diffusion (left panel)
during the accessible time scale, provided the spatial decay of interaction strength is sufficiently slow.
We provide a simple analytical explanation of the mechanism responsible for this quasi-localization, showing that the emergent behavior of the non-equilibrium profiles of local observables around the initial domain-wall positions turns out to be sensitive to the size of the entire system, due to the long range of interactions. 

\emph{Model and protocol}  ---
We consider a ferromagnetic quantum Ising chain with algebraically-decaying couplings, described by the Hamiltonian 
\beq
\label{eq:H}
H= - \frac{J_0}{\mathcal{N}_{\alpha,L}} \sum_{1\le i < j \le L} \frac{\sigma^x_i \sigma^x_j}{\abs{i-j}^{\alpha}} - h \sum_{i=1}^L \sigma^z_i ,
\eeq
where $\sigma^{x,y,z}_i$ are Pauli matrices, $L$ is the number of quantum spins-$1/2$ 
along the chain, $h$ is a transverse magnetic field, and the exponent $\alpha\ge0$ characterizes the range of the ferromagnetic spin-spin interaction. Open boundary conditions are assumed. The rescaling factor $\mathcal{N}_{\alpha,L}=\sum_{1\le i < j \le L} \abs{i-j}^{-\alpha}/(L-1)$ ensures that taking the 
local energy scale $J_0>0$ independent of  $\alpha$ and of the system size results in a proper thermodynamic limit \cite{KacNormalization}.

The behavior of the system described by Eq.~\eqref{eq:H} crucially depends on the value of $\alpha$. 
In the limit $\alpha\to\infty$, Eq.~\eqref{eq:H} reduces to the well-known transverse field quantum Ising chain, which can be solved analytically in terms of free fermions \cite{LSM}. 
The behavior of the system for large enough $\alpha$ may be qualitatively understood in terms of perturbations of this solvable limit (see, e.g., Refs. \cite{SchmidtLRexcitations,VerstraeteLRexcitations}).
In particular, for a sufficiently small transverse field $\abs{h} < h_{\text{cr}}(\alpha)$ the ground state has ferromagnetic order, and elementary excitations above it are essentially dressed domain-walls traveling at a momentum-dependent and bounded speed.
When the interaction decay exponent $\alpha$ is smaller than $2$, however,
ferromagnetic order 
persists at low but finite temperature 
for  $\abs{h}$  smaller than the critical value $h_{\text{cr}}(\alpha)$. 
In this regime, the system can host a broad spectrum of exotic phenomena such as non-linear light-cone propagation of correlations~\cite{HaukeTagliacozzo,Gorshkov2,DaleyEssler}, dynamical phase transitions~\cite{Zunkovic2016,HalimehDPT}, dynamically stabilized Kapitza phases~\cite{LeroseKapitza} and time-crystalline behavior~\cite{NayakPrethermalTC}, some of which have been investigated in a range of experiments 
with trapped ions~\cite{RichermeTrappedIons,JurcevicTrappedIons,ZhangDPT,JurcevicDQPT}.

In this work, we study the non-equilibrium evolution governed by $H$ in Eq.~\eqref{eq:H} starting from initial states $\Ket{\Psi_0} = \Ket{n}$ with a \emph{longitudinal} domain-wall between sites $n$ and $n+1$, i.e., 
\beq
\label{eq:psi0}
\Ket{n} = \bigotimes_{i=1}^{n}\Ket{\uparrow}_i\bigotimes_{i=n+1}^L\Ket{\downarrow}_i \, \equiv \, \Ket{\uparrow_1\dots\uparrow_n \, \downarrow_{n+1}\dots\downarrow_L},
\eeq
where $\Ket{\uparrow,\downarrow}_i$ denotes the eigenstates of $\sigma^x_i$ with eigenvalues $\pm 1$, respectively.
More general initial states $\Ket{n_1,n_2,\dots}$ with multiple domain-walls at positions $n_1,n_2,\dots$ have also been considered. All these states are eigenstates of the Hamiltonian~\eqref{eq:H} for $h=0$. Upon quenching to a non-vanishing field $h \ne 0$, the system undergoes a non-equilibrium evolution, which can be studied by monitoring the dynamics of the spatial profiles of relevant local observables such as, e.g., the magnetization density $\Braket{\sigma^x_i(t)}$.

For large $\alpha$,  
the spatially-localized domain-wall in the initial state represents a superposition of elementary excitations at all possible momenta, and, accordingly, it is expected to exhibit unbounded spreading in space. 
On the other hand, for $\alpha\le2$ isolated domain-walls become highly-excited states, and the spatial propagation of elementary excitations is known to display peculiar features due to the lack of a maximal speed \cite{HaukeTagliacozzo}. In this case, we find a markedly different domain-wall dynamics characterized by spatial localization, as we now detail. 

\begin{figure*}[t]
\centering
\begin{tabular}{cc}
\includegraphics[width=0.45\textwidth]{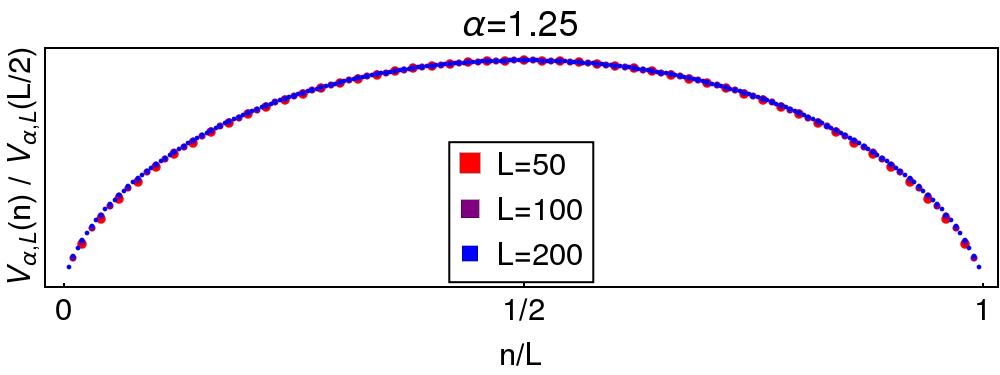} & \includegraphics[width=0.45\textwidth]{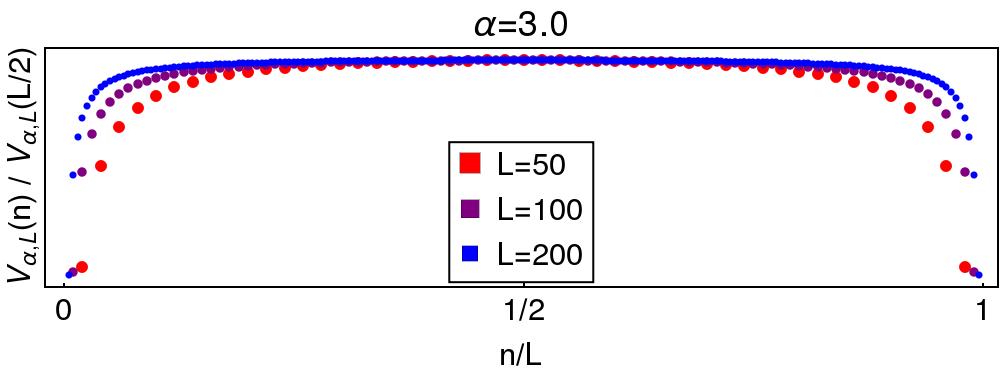}   \\
 \includegraphics[width=0.45\textwidth]{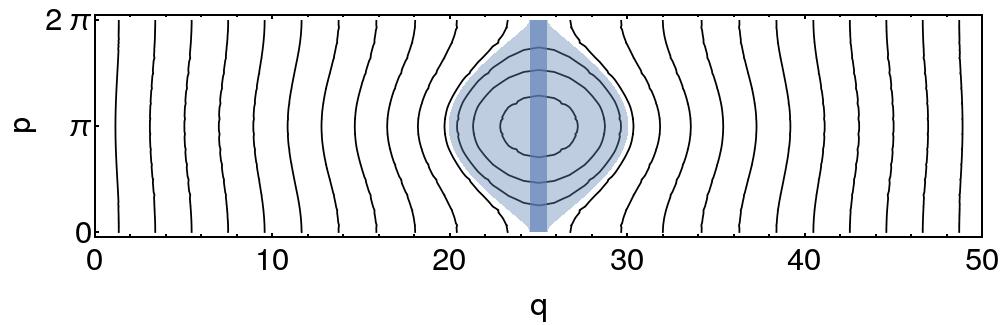} 
 & \includegraphics[width=0.45\textwidth]{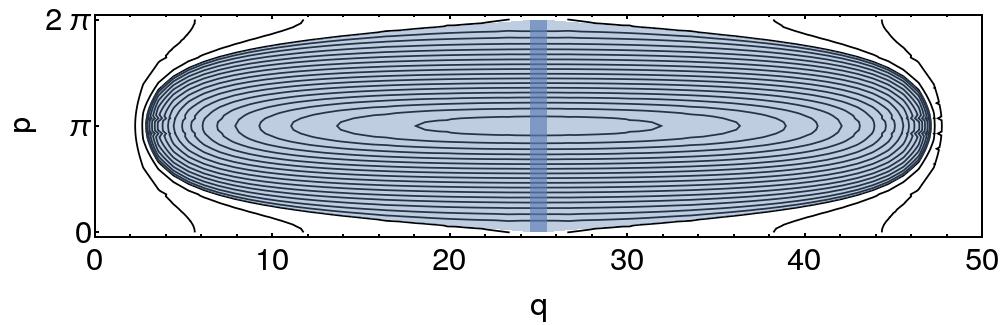}  \\
\end{tabular}
\caption{
Effective description of the dynamics of a single domain-wall.
First row: effective potential $V_{\alpha,L}(n)$ experienced by the domain-wall at position $n$ along the chain,  rescaled by its maximal value $V_{\alpha,L}(L/2)$, as a function of the rescaled lattice position $n/L$, for system sizes $L=50,100,200$.
Second row: sequence of trajectories defined by the classical Hamiltonian \eqref{eq:Hcl}, corresponding to the classical limit of eigenstates of the single-domain-wall problem defined by the Hamiltonian in Eq.~\eqref{eq:Heff} with $h/J_0=0.1$ and $L=50$, obtained from the Bohr-Sommerfeld quantization rule (for a better visualization, one every two possible quantized trajectories is drawn). The coordinate $q \in [1,L-1]$ represents the position along the chain, and the conjugated momentum $p\in[0,2\pi]$ represents the lattice quasi-momentum.
The blue shading illustrates the semiclassical motion of the domain-wall: the initial dark-shaded wavepacket, spatially localized at position $L/2$ and widespread along the momentum axis, traverses the light-shaded region of phase space during its time-evolution. 
}
\label{fig:semiclassical}
\end{figure*}

\emph{Quasi-localization of domain-walls} ---
We perform numerical computations of the non-equilibrium evolution for a range of values of the post-quench transverse field $h/J_0 \in [0,0.5]$ and of the exponent $\alpha \in [0,3]$, and for system sizes $L=50$, $100$, $200$.
We use the second order integrator of the time-dependent variational principle on matrix product states (MPS-TDVP) developed in Refs.~\cite{haegeman2016unifying, haegeman2011time} with the time step $J_0 \, \Delta t = 0.01 $ and bond dimension $D=64$. 
The qualitative features of the results of the simulations appear to depend crucially on $\alpha$ being smaller or larger than $2$.
In particular, as shown in Fig.~\ref{fig:stmagn} for the local magnetization, for $\alpha>2$ unbounded light-cone spreading of the domain-wall occurs from the region around its initial position across the entire system. However, in the presence of longer-range interactions with  $\alpha\le2$, the inhomogeneity initially spreading out from the center of the chain bounces back at a particular characteristic length scale and remains subsequently trapped within this finite region. The spatial amplitude and the temporal duration of this bounce appear to depend on $\alpha$, $h$, and, counterintuitively, on the system size $L$.
This quasi-localization scenario for magnetic defects generalizes to initial states with more domain-walls, as long as their initial separation is larger than the amplitude of their spreading. 
The stability of the magnetization profiles away from the initial positions of domain-walls implies that the lumps of excess energy density, initially concentrated around those positions, remain trapped within the surrounding portions of the chain for a considerable time, highlighting the dramatic slow-down of thermalization and transport.

\emph{Effective single-domain-wall description} ---
We aim now at understanding the mechanism underlying the quasi-localization of domain-walls in the presence of long-range interactions and developing an analytical description of its quantitative features. 
Our analysis is based on the observation that spin-flip processes occurring far away from the domain-wall location and generated by a small transverse field involve, unlike those occurring next to it, a sizeable configurational energy cost as compared to the perturbation strength $h$.
The former processes give rise to small fluctuations of the order parameter away from the domain-wall, and, crucially, the excited ``mesonic'' quasi-particles, made out of two tightly bound domain-walls, have vanishing center-of-mass momentum to lowest order in perturbation theory~\cite{KCTC,GorshkovConfinement,MazzaTransport}. The latter processes, instead, generate effective dynamics of the domain-wall.

We capture these effective dynamics by projecting the many-body quantum Hamiltonian in Eq.~\eqref{eq:H} onto the subspace spanned by the states $\{\Ket{n}\}$ of Eq.~\eqref{eq:psi0} with a single domain-wall located between sites $n$ and $n+1$, with $n=1,2,\dots,L-1$.
Within this approach, which proved to be successful for studying quantum confinement \cite{MazzaTransport},
the matrix elements of the Hamiltonian~\eqref{eq:H} in this subspace may be written as $\braket{n|H|m} = E_{\text{GS}} \delta_{n,m} + \left[ H_{\text{eff}} \right]_{n,m}$, where $E_{GS}=-J_0 (L-1)$ is the ferromagnetic ground state energy, and
\beq
\label{eq:Heff}
\left[ H_{\text{eff}} \right]_{n,m} =  V_{\alpha,L}(n)  \delta_{n,m} - h (\delta_{n,m+1}+\delta_{n,m-1}),
\eeq
where the diagonal  term $V_{\alpha,L}(n)$   is given by the ferromagnetic configurational excess energy 
\beq
\label{eq:V}
V_{\alpha,L}(n) = \frac{2}{\mathcal{N}_{\alpha,L}}\sum_{1\le i \le n} \; \sum_{n+1\le j \le L} \frac{J_0}{\abs{i-j}^{\alpha}}.
\eeq
The quantum evolution starting from a domain-wall $\Ket{n_0}$ in Eq.~\eqref{eq:psi0} is thus approximated by the motion of a single quantum particle  initially placed at the site $n_0$ of a one-dimensional lattice of length $L-1$, hopping to neighboring sites and subject to the potential $V_{\alpha,L}(n)$. 

The resulting qualitative behavior is determined primarily by the shape of the potential $V_{\alpha,L}(n)$, which we display in Fig.~\ref{fig:semiclassical} for some representative values of $\alpha$. 
One realizes that for $\alpha>2$, the potential  becomes flat (i.e., independent of $n$) in the limit $L\to\infty$, and hence eigenstates approach spatially-extended plane waves characterized by their momentum. Conversely, for $1<\alpha\le 2$ the spatial dependence of $V_{\alpha,L}(n)$ has a non-trivial scaling limit for large $L$, described by a smooth function $\mathcal{V}_\alpha$:
\beq
\label{eq:Vscaling}
V_{\alpha,L}(n) \; \underset{L\to\infty}{\thicksim}\; c_\alpha J_0   L^{2-\alpha} \; \mathcal{V}_\alpha \big(n/L \big) ,
\eeq
with $c_\alpha = 2/ [(2-\alpha)(\alpha-1) \zeta(\alpha)]$, $\zeta$ the Riemann zeta-function and $\mathcal{V}_\alpha(x) = x^{2-\alpha} + (1-x)^{2-\alpha}-1$,
which is obtained by estimating the sums in Eq.~\eqref{eq:V}. (In the limiting case $\alpha=2$, powers are substituted by logarithms.)
In particular, due to the concurrence of unbounded potential energy $\sim J_0 L^{2-\alpha}$ and bounded kinetic energy $\sim h$ on the lattice, energy conservation implies that a particle initially placed at a given lattice site can travel at most a finite distance away from it. Correspondingly, all eigenstates of $H_{\text{eff}}$ are spatially localized \footnote{At finite $L$, due to inversion symmetry with respect to the center of the open chain, the eigenstates of $H_{\text{eff}}$ are actually even and odd superpositions of wavefunctions localized at symmetric positions in the chain. However, such a finite-size effect is unstable to any perturbation which breaks this symmetry, such as, e.g., random boundary conditions.}.
This sharp transition in the structure of the spectrum of $H_{\text{eff}}$ in Eq.~\eqref{eq:Heff} upon decreasing $\alpha$ below $2$ is illustrated in Fig.~\ref{fig:semiclassical} and provides a simple explanation for the onset of quasi-localization in the presence of long-range interactions, as shown in Fig.~\ref{fig:stmagn}.

\begin{figure*}[t]
\begin{tabular}{cc}
\includegraphics[width=0.45\textwidth]{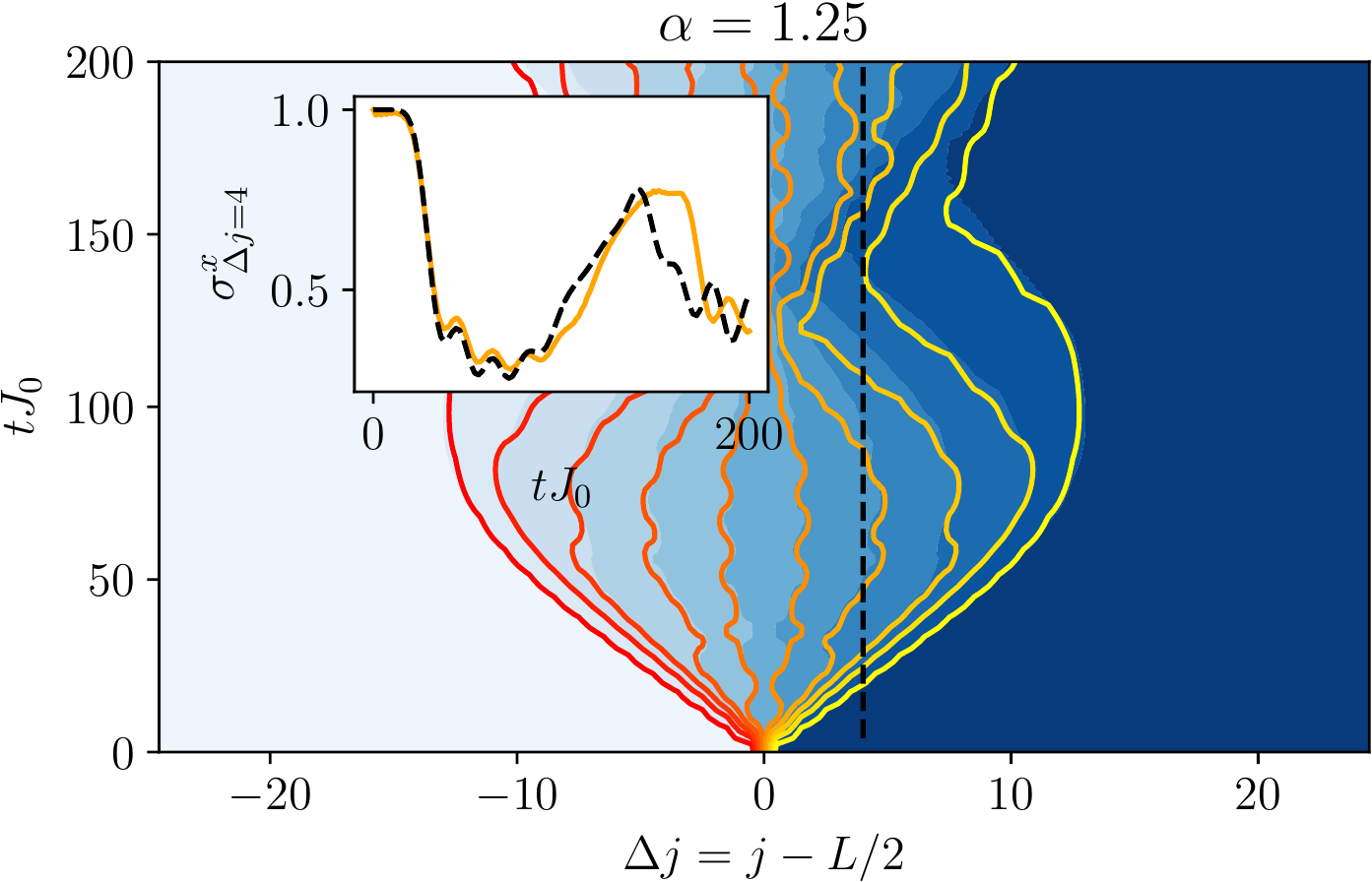} & \includegraphics[width=0.45\textwidth]{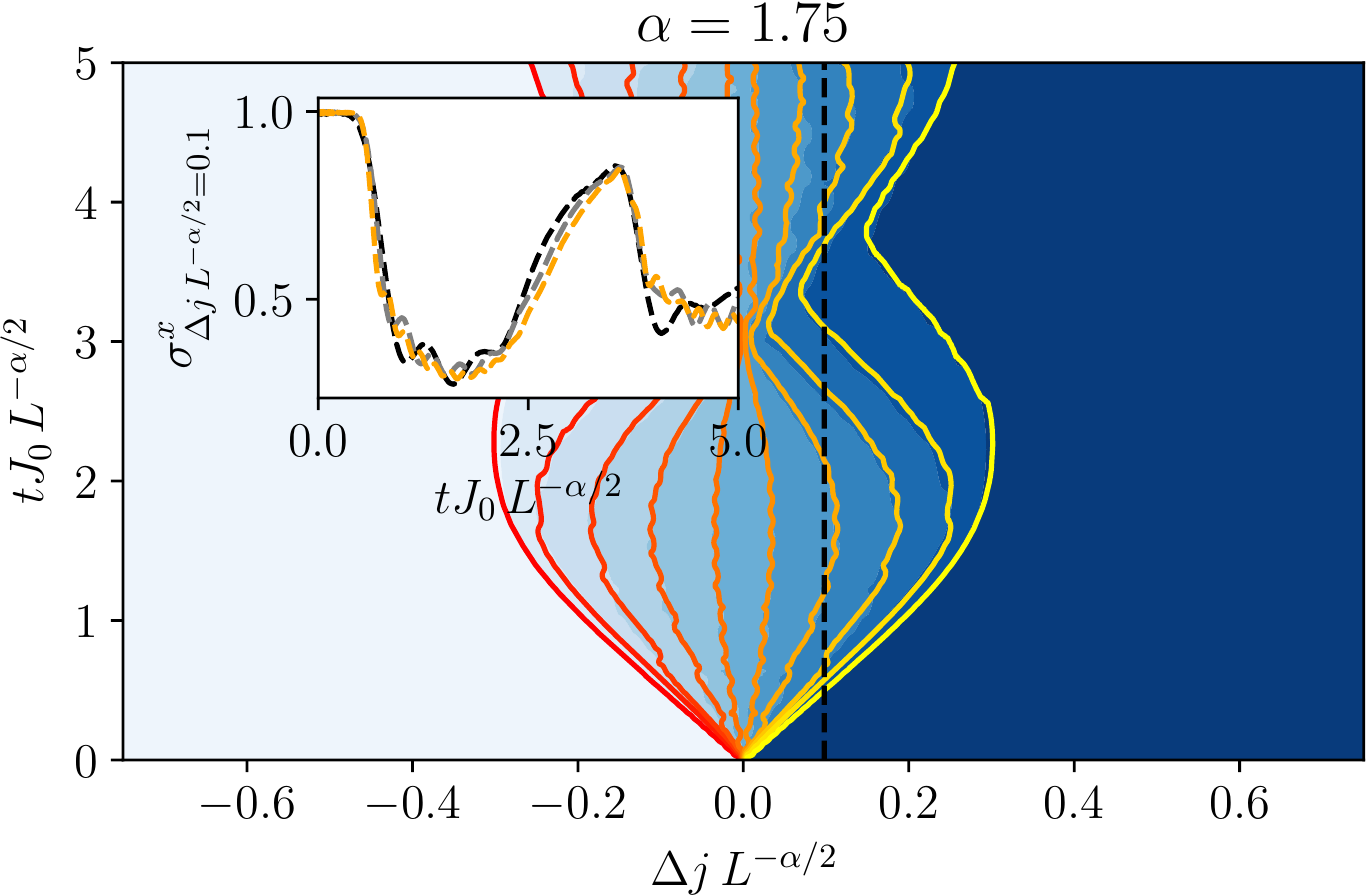}
\end{tabular}
\caption{
Left: Quantitative comparison between the non-equilibrium evolution of the magnetization obtained from MPS-TDVP numerical simulations 
(blue scale, background coloring) and those predicted by the single-domain-wall approximation in Eq.~\eqref{eq:mprofileapprox} (yellow-red scale, solid contour lines). These data refer to $\alpha=1.25$, $h/J_0=0.1$, $L=200$. Inset: time slice of the plot at the indicated position along the chain, with a comparison between the MPS-TDVP data (dashed black line) and the result of the approximation (solid orange).
Right: Quantitative verification of the validity of the semiclassical scaling laws of the profiles in Eq.~\eqref{eq:scalingprofilesat} with respect to the system size $L$. These data are obtained from MPS-TDVP simulations with $\alpha=1.75$, $h/J_0=0.1$, and $L=100$ (blue scale, background coloring) or $200$ (yellow-red scale, solid contour lines).
Inset: time slice of the plot at the indicated rescaled position along the chain, with a comparison between $L=100$ (dashed black line) and $L=200$ (dashed orange).
In both panels, data are converged with bond dimension $D=64$. 
}
\label{fig:comparison}
\end{figure*}



Note that the quasi-localization of magnetic defects discussed above, takes place only if the effective potential $V_{\alpha,L}$ is confining at large distances, i.e., for $\alpha\le2$. Accordingly, this phenomenon it is not directly related with the 
confined dynamics of correlations functions \cite{GorshkovConfinement} and anomalous cusps in the time-evolution of the Loschmidt echo 
\cite{Halimeh:ADQPTandConfinement}, which occur whenever the effective potential 
possesses 
bound states, i.e., also for $\alpha>2$ in the present model.

For $0\le\alpha\le1$ and in the thermodynamic limit, the ferromagnetic interaction is exactly described by its mean-field approximation
\cite{mori2018prethermalization} and thus it is equivalent to that between a spin and 
the self-consistent longitudinal field 
$\lambda_j(t)=\frac{J_0}{2\mathcal{N}_{\alpha,L}}\sum_{i(\ne j)}^L \frac{\langle \sigma^x_i (t) \rangle}{|i-j|^\alpha}$, which can actually prevent domain-walls from spreading in space.
This is the reason why, in the following, we focus on the more interesting case $1<\alpha\le 2$ corresponding to intermediate-range interactions.

We test the effectiveness of the above single-domain-wall approximation by quantitatively comparing its results with those of the MPS-TDVP simulations. 
Relevant observables are projected onto the single-domain-wall subspace, and their expectation on the evolving state $\Ket{\widetilde{\Psi}(t)}=e^{-iH_{\text{eff}}t} \Ket{n_0}$ is computed. We primarily focus on the local magnetization $m_j(t)=\Braket{\widetilde{\Psi}(t)|\sigma^x_j | \widetilde{\Psi}(t)}$,   
which reads 
\beq
\label{eq:mprofileapprox}
m_j(t) = 1-2\sum_{1\le n<j} \big\lvert \braket{n|\widetilde{\Psi}(t)}\big\rvert^2.
\eeq 
The quantitative comparison between this profile and the corresponding one obtained from MPS-TDVP simulations is shown in the left panel of Fig.~\ref{fig:comparison} with solid contour lines and background coloring, respectively, and in the relative inset for the time-evolving magnetization at a fixed position along the chain  indicated by the vertical black dashed line. The agreement turns out to be excellent for small $h$ and up to considerably long times. It appears to slowly deteriorate at long times, with an associated time scale that decreases upon increasing $h$. However, in all cases, the qualitative agreement remains fairly good for all accessible times.

\emph{Semiclassical scaling laws} ---
The above single-domain-wall approach allows one to derive the quantitative features of the quasi-localization of domain-walls and, in particular, to elucidate the role of long-range interactions in determining its strong sensitivity to the system size. This is conveniently achieved through a semiclassical treatment of the single-particle quantum-mechanical problem defined by the Hamiltonian in Eq.~\eqref{eq:Heff}, which describes its solution with increasing accuracy as $L\to\infty$ \cite{RutkevichSemiclassical}.

Taking the classical limit in Eqs.~\eqref{eq:Heff} and~\eqref{eq:Vscaling} 
defined on the phase space $(q,p) \in [1,L-1]\times[0,2\pi]$ with periodic boundary conditions on the momentum $p$, one finds for $\alpha>1$
\beq
\label{eq:Hcl}
\mathcal{H}_{\text{cl}}(q,p) = - 2 h \cos p \, + c_\alpha J_0  \Big[ q^{2-\alpha} + (L-q)^{2-\alpha}-L^{2-\alpha} \Big].
\eeq
The semiclassical trajectories are defined by Hamilton's equations $\dot{q} = \partial_p \mathcal{H}_{\text{cl}}$ and $\dot{p} = - \partial_q \mathcal{H}_{\text{cl}}$.
Fig.~\ref{fig:semiclassical} shows the structure of the semiclassical eigenstates for $1<\alpha\le 2$ (left) and $\alpha>2$ (right), obtained by quantizing the phase space area encircled by classical trajectories, illustrating the quasi-localization transition.

Scaling laws for the motion of quasi-localized domain-walls as a function of the parameters and of the system size can be derived within this semiclassical treatment. In particular, by evaluating the extension and frequency of the classical trajectories crossing a position $n_0$ along the chain (see Fig.~\ref{fig:semiclassical} for an illustration), one readily derives an estimate of the spatial width $\Delta j$  and of the temporal period $\Delta t$ of the spatiotemporal profiles, given by
\beq
\label{eq:scalingprofilesaway}
\Delta j \propto h L^{\alpha-1}\; , \qquad \Delta t \propto L^{\alpha-1}\; ,
\eeq
respectively, for domain-walls far away from the center of the chain, i.e., with $n_0 \ll L/2$ or $n_0 \gg L/2$, and
\beq
\label{eq:scalingprofilesat}
\Delta j \propto \sqrt{h L^{\alpha}}\; , \qquad \Delta t \propto \sqrt{L^{\alpha} / h}\; ,
\eeq
respectively, for domain-walls near the center of the chain, i.e., with $n_0 \approx L/2$.
These predicted scaling laws are confirmed by the non-equilibrium profiles obtained via MPS-TDVP to a high degree of accuracy, as shown, e.g., in the right panel of Fig.~\ref{fig:comparison} with rescaled numerical data for various system sizes $L$ superimposed. Thus, the counterintuitive dependence of the local dynamical behavior of confined domain-walls on the chain length $L$, observed in the numerical simulations, is actually determined by the system-size scaling of the configurational energy $V_{\alpha,L}(n)$ approximately governing the evolution of isolated domain-walls, which is, in turn, a manifestation of the long-range nature of the interactions.

\emph{Conclusion and outlook} ---
In this work, we have shown that a non-equilibrium transition occurs in the dynamics of variable-range quantum Ising chains with magnetic defects, characterized by a sharp change from an unbounded spatial spreading of domain-walls to quasi-localized dynamics upon increasing the range of interactions.
We have provided an analytical understanding of this occurrence using an approximate description
obtained by neglecting multi-domain-wall processes. This approach accurately reproduces the results of the MPS-TDVP simulations up to the simulation times explored in the present study. The phenomena reported here can be readily tested in current experiments with trapped ions.

Long-range interacting systems are known to exhibit slow relaxation~\cite{NeyenhuisPrethermalizationLRExp,mori2018prethermalization} and slow entanglement growth~\cite{Daley,DaleyEssler,GorshkovConfinement,LeroseEE}, reminiscent of many-body localized dynamics. 
The quasi-localization of excitations reported in this work suggests a further qualitative analogy along these lines.
Due to the absence of strictly conserved localized quantities, though, the reported scenario is expected to be superseded at long times by processes leading to thermalization. However, a proper investigation of this issue requires a refined analysis of spectral properties~\cite{KonikConfinement} or of high orders in perturbation theory~ \cite{DeRoeck1,AbaninProsen:SlowDynamics,LinMotrunich:Quasiconserved}, and is left for future investigations.

\emph{Acknowledgements} --- We thank G. Pagano for discussing with us the experimental feasibility of our setup with trapped ions. B. \v{Z}. acknowledges support  by the Advanced grant of European Research Council (ERC), No. 694544 – OMNES.

\bibliography{biblio2} 

\begin{thebibliography}{53}%
\makeatletter
\providecommand \@ifxundefined [1]{%
 \@ifx{#1\undefined}
}%
\providecommand \@ifnum [1]{%
 \ifnum #1\expandafter \@firstoftwo
 \else \expandafter \@secondoftwo
 \fi
}%
\providecommand \@ifx [1]{%
 \ifx #1\expandafter \@firstoftwo
 \else \expandafter \@secondoftwo
 \fi
}%
\providecommand \natexlab [1]{#1}%
\providecommand \enquote  [1]{``#1''}%
\providecommand \bibnamefont  [1]{#1}%
\providecommand \bibfnamefont [1]{#1}%
\providecommand \citenamefont [1]{#1}%
\providecommand \href@noop [0]{\@secondoftwo}%
\providecommand \href [0]{\begingroup \@sanitize@url \@href}%
\providecommand \@href[1]{\@@startlink{#1}\@@href}%
\providecommand \@@href[1]{\endgroup#1\@@endlink}%
\providecommand \@sanitize@url [0]{\catcode `\\12\catcode `\$12\catcode
  `\&12\catcode `\#12\catcode `\^12\catcode `\_12\catcode `\%12\relax}%
\providecommand \@@startlink[1]{}%
\providecommand \@@endlink[0]{}%
\providecommand \url  [0]{\begingroup\@sanitize@url \@url }%
\providecommand \@url [1]{\endgroup\@href {#1}{\urlprefix }}%
\providecommand \urlprefix  [0]{URL }%
\providecommand \Eprint [0]{\href }%
\providecommand \doibase [0]{http://dx.doi.org/}%
\providecommand \selectlanguage [0]{\@gobble}%
\providecommand \bibinfo  [0]{\@secondoftwo}%
\providecommand \bibfield  [0]{\@secondoftwo}%
\providecommand \translation [1]{[#1]}%
\providecommand \BibitemOpen [0]{}%
\providecommand \bibitemStop [0]{}%
\providecommand \bibitemNoStop [0]{.\EOS\space}%
\providecommand \EOS [0]{\spacefactor3000\relax}%
\providecommand \BibitemShut  [1]{\csname bibitem#1\endcsname}%
\let\auto@bib@innerbib\@empty
\bibitem [{\citenamefont {Datta}(1995)}]{mesoscopic1}%
  \BibitemOpen
  \bibfield  {author} {\bibinfo {author} {\bibfnamefont {S.}~\bibnamefont
  {Datta}},\ }\href@noop {} {\emph {\bibinfo {title} {Electronic transport in
  mesoscopic systems}}}\ (\bibinfo  {publisher} {Cambridge University Press},\
  \bibinfo {year} {1995})\BibitemShut {NoStop}%
\bibitem [{\citenamefont {Akkermans}\ and\ \citenamefont
  {Montambaux}(2007)}]{mesoscopic2}%
  \BibitemOpen
  \bibfield  {author} {\bibinfo {author} {\bibfnamefont {E.}~\bibnamefont
  {Akkermans}}\ and\ \bibinfo {author} {\bibfnamefont {G.}~\bibnamefont
  {Montambaux}},\ }\href@noop {} {\emph {\bibinfo {title} {Mesoscopic Physics
  of Electrons and Photons}}}\ (\bibinfo  {publisher} {Cambridge University
  Press},\ \bibinfo {year} {2007})\BibitemShut {NoStop}%
\bibitem [{\citenamefont {Calabrese}\ and\ \citenamefont
  {Cardy}(2005)}]{CC_QuasiparticlePicture}%
  \BibitemOpen
  \bibfield  {author} {\bibinfo {author} {\bibfnamefont {P.}~\bibnamefont
  {Calabrese}}\ and\ \bibinfo {author} {\bibfnamefont {J.}~\bibnamefont
  {Cardy}},\ }\href {http://stacks.iop.org/1742-5468/2005/i=04/a=P04010}
  {\bibfield  {journal} {\bibinfo  {journal} {Journal of Statistical Mechanics:
  Theory and Experiment}\ }\textbf {\bibinfo {volume} {2005}},\ \bibinfo
  {pages} {P04010} (\bibinfo {year} {2005})}\BibitemShut {NoStop}%
\bibitem [{\citenamefont {Kim}\ and\ \citenamefont {Huse}(2013)}]{KimHuse}%
  \BibitemOpen
  \bibfield  {author} {\bibinfo {author} {\bibfnamefont {H.}~\bibnamefont
  {Kim}}\ and\ \bibinfo {author} {\bibfnamefont {D.~A.}\ \bibnamefont {Huse}},\
  }\href {\doibase 10.1103/PhysRevLett.111.127205} {\bibfield  {journal}
  {\bibinfo  {journal} {Phys. Rev. Lett.}\ }\textbf {\bibinfo {volume} {111}},\
  \bibinfo {pages} {127205} (\bibinfo {year} {2013})}\BibitemShut {NoStop}%
\bibitem [{\citenamefont {Anderson}(1958)}]{AndersonLocalization}%
  \BibitemOpen
  \bibfield  {author} {\bibinfo {author} {\bibfnamefont {P.~W.}\ \bibnamefont
  {Anderson}},\ }\href {\doibase 10.1103/PhysRev.109.1492} {\bibfield
  {journal} {\bibinfo  {journal} {Phys. Rev.}\ }\textbf {\bibinfo {volume}
  {109}},\ \bibinfo {pages} {1492} (\bibinfo {year} {1958})}\BibitemShut
  {NoStop}%
\bibitem [{\citenamefont {Basko}\ \emph {et~al.}(2006)\citenamefont {Basko},
  \citenamefont {Aleiner},\ and\ \citenamefont {Altshuler}}]{BAA}%
  \BibitemOpen
  \bibfield  {author} {\bibinfo {author} {\bibfnamefont {D.}~\bibnamefont
  {Basko}}, \bibinfo {author} {\bibfnamefont {I.}~\bibnamefont {Aleiner}}, \
  and\ \bibinfo {author} {\bibfnamefont {B.}~\bibnamefont {Altshuler}},\ }\href
  {\doibase https://doi.org/10.1016/j.aop.2005.11.014} {\bibfield  {journal}
  {\bibinfo  {journal} {Annals of Physics}\ }\textbf {\bibinfo {volume}
  {321}},\ \bibinfo {pages} {1126 } (\bibinfo {year} {2006})}\BibitemShut
  {NoStop}%
\bibitem [{\citenamefont {Huse}\ \emph {et~al.}(2014)\citenamefont {Huse},
  \citenamefont {Nandkishore},\ and\ \citenamefont
  {Oganesyan}}]{HuseNandkishore}%
  \BibitemOpen
  \bibfield  {author} {\bibinfo {author} {\bibfnamefont {D.~A.}\ \bibnamefont
  {Huse}}, \bibinfo {author} {\bibfnamefont {R.}~\bibnamefont {Nandkishore}}, \
  and\ \bibinfo {author} {\bibfnamefont {V.}~\bibnamefont {Oganesyan}},\ }\href
  {\doibase 10.1103/PhysRevB.90.174202} {\bibfield  {journal} {\bibinfo
  {journal} {Phys. Rev. B}\ }\textbf {\bibinfo {volume} {90}},\ \bibinfo
  {pages} {174202} (\bibinfo {year} {2014})}\BibitemShut {NoStop}%
\bibitem [{\citenamefont {Altman}(2018)}]{AltmanMBL}%
  \BibitemOpen
  \bibfield  {author} {\bibinfo {author} {\bibfnamefont {E.}~\bibnamefont
  {Altman}},\ }\href@noop {} {\bibfield  {journal} {\bibinfo  {journal} {Nature
  Physics}\ }\textbf {\bibinfo {volume} {14}},\ \bibinfo {pages} {979}
  (\bibinfo {year} {2018})}\BibitemShut {NoStop}%
\bibitem [{\citenamefont {\ifmmode \check{Z}\else
  \v{Z}\fi{}nidari\ifmmode~\check{c}\else \v{c}\fi{}}\ \emph
  {et~al.}(2008)\citenamefont {\ifmmode \check{Z}\else
  \v{Z}\fi{}nidari\ifmmode~\check{c}\else \v{c}\fi{}}, \citenamefont {Prosen},\
  and\ \citenamefont {Prelov\ifmmode~\check{s}\else \v{s}\fi{}ek}}]{ProsenMBL}%
  \BibitemOpen
  \bibfield  {author} {\bibinfo {author} {\bibfnamefont {M.}~\bibnamefont
  {\ifmmode \check{Z}\else \v{Z}\fi{}nidari\ifmmode~\check{c}\else
  \v{c}\fi{}}}, \bibinfo {author} {\bibfnamefont {T.~c.~v.}\ \bibnamefont
  {Prosen}}, \ and\ \bibinfo {author} {\bibfnamefont {P.}~\bibnamefont
  {Prelov\ifmmode~\check{s}\else \v{s}\fi{}ek}},\ }\href {\doibase
  10.1103/PhysRevB.77.064426} {\bibfield  {journal} {\bibinfo  {journal} {Phys.
  Rev. B}\ }\textbf {\bibinfo {volume} {77}},\ \bibinfo {pages} {064426}
  (\bibinfo {year} {2008})}\BibitemShut {NoStop}%
\bibitem [{\citenamefont {Bardarson}\ \emph {et~al.}(2012)\citenamefont
  {Bardarson}, \citenamefont {Pollmann},\ and\ \citenamefont
  {Moore}}]{BardarsonPollmanMoore}%
  \BibitemOpen
  \bibfield  {author} {\bibinfo {author} {\bibfnamefont {J.~H.}\ \bibnamefont
  {Bardarson}}, \bibinfo {author} {\bibfnamefont {F.}~\bibnamefont {Pollmann}},
  \ and\ \bibinfo {author} {\bibfnamefont {J.~E.}\ \bibnamefont {Moore}},\
  }\href {\doibase 10.1103/PhysRevLett.109.017202} {\bibfield  {journal}
  {\bibinfo  {journal} {Phys. Rev. Lett.}\ }\textbf {\bibinfo {volume} {109}},\
  \bibinfo {pages} {017202} (\bibinfo {year} {2012})}\BibitemShut {NoStop}%
\bibitem [{\citenamefont {Vasseur}\ and\ \citenamefont
  {Moore}(2016)}]{VasseurMooreReview}%
  \BibitemOpen
  \bibfield  {author} {\bibinfo {author} {\bibfnamefont {R.}~\bibnamefont
  {Vasseur}}\ and\ \bibinfo {author} {\bibfnamefont {J.~E.}\ \bibnamefont
  {Moore}},\ }\href {http://stacks.iop.org/1742-5468/2016/i=6/a=064010}
  {\bibfield  {journal} {\bibinfo  {journal} {Journal of Statistical Mechanics:
  Theory and Experiment}\ }\textbf {\bibinfo {volume} {2016}},\ \bibinfo
  {pages} {064010} (\bibinfo {year} {2016})}\BibitemShut {NoStop}%
\bibitem [{\citenamefont {De~Roeck}\ and\ \citenamefont
  {Huveneers}(2014)}]{DeRoeck1}%
  \BibitemOpen
  \bibfield  {author} {\bibinfo {author} {\bibfnamefont {W.}~\bibnamefont
  {De~Roeck}}\ and\ \bibinfo {author} {\bibfnamefont {F.}~\bibnamefont
  {Huveneers}},\ }\href {https://doi.org/10.1007/s00220-014-2116-8} {\bibfield
  {journal} {\bibinfo  {journal} {Commun. Math. Phys.}\ }\textbf {\bibinfo
  {volume} {332}},\ \bibinfo {pages} {1017} (\bibinfo {year}
  {2014})}\BibitemShut {NoStop}%
\bibitem [{\citenamefont {Schiulaz}\ \emph {et~al.}(2015)\citenamefont
  {Schiulaz}, \citenamefont {Silva},\ and\ \citenamefont
  {M\"uller}}]{Schiulaz1}%
  \BibitemOpen
  \bibfield  {author} {\bibinfo {author} {\bibfnamefont {M.}~\bibnamefont
  {Schiulaz}}, \bibinfo {author} {\bibfnamefont {A.}~\bibnamefont {Silva}}, \
  and\ \bibinfo {author} {\bibfnamefont {M.}~\bibnamefont {M\"uller}},\ }\href
  {\doibase 10.1103/PhysRevB.91.184202} {\bibfield  {journal} {\bibinfo
  {journal} {Phys. Rev. B}\ }\textbf {\bibinfo {volume} {91}},\ \bibinfo
  {pages} {184202} (\bibinfo {year} {2015})}\BibitemShut {NoStop}%
\bibitem [{\citenamefont {Yao}\ \emph {et~al.}(2016)\citenamefont {Yao},
  \citenamefont {Laumann}, \citenamefont {Cirac}, \citenamefont {Lukin},\ and\
  \citenamefont {Moore}}]{Cirac:QuasiMBLWithoutDisorder}%
  \BibitemOpen
  \bibfield  {author} {\bibinfo {author} {\bibfnamefont {N.~Y.}\ \bibnamefont
  {Yao}}, \bibinfo {author} {\bibfnamefont {C.~R.}\ \bibnamefont {Laumann}},
  \bibinfo {author} {\bibfnamefont {J.~I.}\ \bibnamefont {Cirac}}, \bibinfo
  {author} {\bibfnamefont {M.~D.}\ \bibnamefont {Lukin}}, \ and\ \bibinfo
  {author} {\bibfnamefont {J.~E.}\ \bibnamefont {Moore}},\ }\href {\doibase
  10.1103/PhysRevLett.117.240601} {\bibfield  {journal} {\bibinfo  {journal}
  {Phys. Rev. Lett.}\ }\textbf {\bibinfo {volume} {117}},\ \bibinfo {pages}
  {240601} (\bibinfo {year} {2016})}\BibitemShut {NoStop}%
\bibitem [{\citenamefont {Carleo}\ \emph {et~al.}(2012)\citenamefont {Carleo},
  \citenamefont {Becca}, \citenamefont {Schirò},\ and\ \citenamefont
  {Fabrizio}}]{CarleoGlassy}%
  \BibitemOpen
  \bibfield  {author} {\bibinfo {author} {\bibfnamefont {G.}~\bibnamefont
  {Carleo}}, \bibinfo {author} {\bibfnamefont {F.}~\bibnamefont {Becca}},
  \bibinfo {author} {\bibfnamefont {M.}~\bibnamefont {Schirò}}, \ and\
  \bibinfo {author} {\bibfnamefont {M.}~\bibnamefont {Fabrizio}},\ }\href
  {http://dx.doi.org/10.1038/srep00243} {\bibfield  {journal} {\bibinfo
  {journal} {Scientific Reports}\ }\textbf {\bibinfo {volume} {2}},\ \bibinfo
  {pages} {243} (\bibinfo {year} {2012})}\BibitemShut {NoStop}%
\bibitem [{\citenamefont {Michailidis}\ \emph {et~al.}(2018)\citenamefont
  {Michailidis}, \citenamefont {\ifmmode \check{Z}\else
  \v{Z}\fi{}nidari\ifmmode~\check{c}\else \v{c}\fi{}}, \citenamefont
  {Medvedyeva}, \citenamefont {Abanin}, \citenamefont {Prosen},\ and\
  \citenamefont {Papi\ifmmode~\acute{c}\else
  \'{c}\fi{}}}]{AbaninProsen:SlowDynamics}%
  \BibitemOpen
  \bibfield  {author} {\bibinfo {author} {\bibfnamefont {A.~A.}\ \bibnamefont
  {Michailidis}}, \bibinfo {author} {\bibfnamefont {M.}~\bibnamefont {\ifmmode
  \check{Z}\else \v{Z}\fi{}nidari\ifmmode~\check{c}\else \v{c}\fi{}}}, \bibinfo
  {author} {\bibfnamefont {M.}~\bibnamefont {Medvedyeva}}, \bibinfo {author}
  {\bibfnamefont {D.~A.}\ \bibnamefont {Abanin}}, \bibinfo {author}
  {\bibfnamefont {T.~c.~v.}\ \bibnamefont {Prosen}}, \ and\ \bibinfo {author}
  {\bibfnamefont {Z.}~\bibnamefont {Papi\ifmmode~\acute{c}\else \'{c}\fi{}}},\
  }\href {\doibase 10.1103/PhysRevB.97.104307} {\bibfield  {journal} {\bibinfo
  {journal} {Phys. Rev. B}\ }\textbf {\bibinfo {volume} {97}},\ \bibinfo
  {pages} {104307} (\bibinfo {year} {2018})}\BibitemShut {NoStop}%
\bibitem [{\citenamefont {Choudhury}\ \emph {et~al.}(2018)\citenamefont
  {Choudhury}, \citenamefont {Kim},\ and\ \citenamefont
  {Zhou}}]{QuasiMBLFrystration}%
  \BibitemOpen
  \bibfield  {author} {\bibinfo {author} {\bibfnamefont {S.}~\bibnamefont
  {Choudhury}}, \bibinfo {author} {\bibfnamefont {E.}~\bibnamefont {Kim}}, \
  and\ \bibinfo {author} {\bibfnamefont {Q.}~\bibnamefont {Zhou}},\ }\href@noop
  {} {\bibfield  {journal} {\bibinfo  {journal} {arXiv:1807.05969}\ } (\bibinfo
  {year} {2018})}\BibitemShut {NoStop}%
\bibitem [{\citenamefont {van Nieuwenburg}\ \emph {et~al.}(2018)\citenamefont
  {van Nieuwenburg}, \citenamefont {Baum},\ and\ \citenamefont
  {Refael}}]{Refael:MBLBlochOscillations}%
  \BibitemOpen
  \bibfield  {author} {\bibinfo {author} {\bibfnamefont {E.~P.~L.}\
  \bibnamefont {van Nieuwenburg}}, \bibinfo {author} {\bibfnamefont
  {Y.}~\bibnamefont {Baum}}, \ and\ \bibinfo {author} {\bibfnamefont
  {G.}~\bibnamefont {Refael}},\ }\href@noop {} {\bibfield  {journal} {\bibinfo
  {journal} {arxiv:1808.00471}\ } (\bibinfo {year} {2018})}\BibitemShut
  {NoStop}%
\bibitem [{\citenamefont {Schulz}\ \emph {et~al.}(2018)\citenamefont {Schulz},
  \citenamefont {Hooley}, \citenamefont {Moessner},\ and\ \citenamefont
  {Pollmann}}]{Pollmann:MBLBlochOscillations}%
  \BibitemOpen
  \bibfield  {author} {\bibinfo {author} {\bibfnamefont {M.}~\bibnamefont
  {Schulz}}, \bibinfo {author} {\bibfnamefont {C.~A.}\ \bibnamefont {Hooley}},
  \bibinfo {author} {\bibfnamefont {R.}~\bibnamefont {Moessner}}, \ and\
  \bibinfo {author} {\bibfnamefont {F.}~\bibnamefont {Pollmann}},\ }\href@noop
  {} {\bibfield  {journal} {\bibinfo  {journal} {arxiv:1808.01250}\ } (\bibinfo
  {year} {2018})}\BibitemShut {NoStop}%
\bibitem [{\citenamefont {Kormos}\ \emph {et~al.}(2016)\citenamefont {Kormos},
  \citenamefont {Collura}, \citenamefont {Takács},\ and\ \citenamefont
  {Calabrese}}]{KCTC}%
  \BibitemOpen
  \bibfield  {author} {\bibinfo {author} {\bibfnamefont {M.}~\bibnamefont
  {Kormos}}, \bibinfo {author} {\bibfnamefont {M.}~\bibnamefont {Collura}},
  \bibinfo {author} {\bibfnamefont {G.}~\bibnamefont {Takács}}, \ and\
  \bibinfo {author} {\bibfnamefont {P.}~\bibnamefont {Calabrese}},\ }\href@noop
  {} {\bibfield  {journal} {\bibinfo  {journal} {Nature Physics}\ }\textbf
  {\bibinfo {volume} {13}},\ \bibinfo {pages} {246} (\bibinfo {year}
  {2016})}\BibitemShut {NoStop}%
\bibitem [{\citenamefont {Mazza}\ \emph {et~al.}(2018)\citenamefont {Mazza},
  \citenamefont {Perfetto}, \citenamefont {Lerose}, \citenamefont {Collura},\
  and\ \citenamefont {Gambassi}}]{MazzaTransport}%
  \BibitemOpen
  \bibfield  {author} {\bibinfo {author} {\bibfnamefont {P.~P.}\ \bibnamefont
  {Mazza}}, \bibinfo {author} {\bibfnamefont {G.}~\bibnamefont {Perfetto}},
  \bibinfo {author} {\bibfnamefont {A.}~\bibnamefont {Lerose}}, \bibinfo
  {author} {\bibfnamefont {M.}~\bibnamefont {Collura}}, \ and\ \bibinfo
  {author} {\bibfnamefont {A.}~\bibnamefont {Gambassi}},\ }\href@noop {}
  {\bibfield  {journal} {\bibinfo  {journal} {arXiv:1806.09674}\ } (\bibinfo
  {year} {2018})}\BibitemShut {NoStop}%
\bibitem [{\citenamefont {Liu}\ \emph {et~al.}(2018)\citenamefont {Liu} \emph
  {et~al.}}]{GorshkovConfinement}%
  \BibitemOpen
  \bibfield  {author} {\bibinfo {author} {\bibfnamefont {F.}~\bibnamefont
  {Liu}} \emph {et~al.},\ }\href@noop {} {\bibfield  {journal} {\bibinfo
  {journal} {arXiv:1810.02365}\ } (\bibinfo {year} {2018})}\BibitemShut
  {NoStop}%
\bibitem [{\citenamefont {Britton}\ \emph {et~al.}(2012)\citenamefont {Britton}
  \emph {et~al.}}]{BrittonTrappedIons}%
  \BibitemOpen
  \bibfield  {author} {\bibinfo {author} {\bibfnamefont {J.~W.}\ \bibnamefont
  {Britton}} \emph {et~al.},\ }\href {http://dx.doi.org/10.1038/nature10981}
  {\bibfield  {journal} {\bibinfo  {journal} {Nature}\ }\textbf {\bibinfo
  {volume} {484}},\ \bibinfo {pages} {489} (\bibinfo {year}
  {2012})}\BibitemShut {NoStop}%
\bibitem [{\citenamefont {Richerme}\ \emph {et~al.}(2014)\citenamefont
  {Richerme} \emph {et~al.}}]{RichermeTrappedIons}%
  \BibitemOpen
  \bibfield  {author} {\bibinfo {author} {\bibfnamefont {P.}~\bibnamefont
  {Richerme}} \emph {et~al.},\ }\href {http://dx.doi.org/10.1038/nature13450}
  {\bibfield  {journal} {\bibinfo  {journal} {Nature}\ }\textbf {\bibinfo
  {volume} {511}},\ \bibinfo {pages} {198} (\bibinfo {year}
  {2014})}\BibitemShut {NoStop}%
\bibitem [{\citenamefont {Jurcevic}\ \emph {et~al.}(2014)\citenamefont
  {Jurcevic} \emph {et~al.}}]{JurcevicTrappedIons}%
  \BibitemOpen
  \bibfield  {author} {\bibinfo {author} {\bibfnamefont {P.}~\bibnamefont
  {Jurcevic}} \emph {et~al.},\ }\href {http://dx.doi.org/10.1038/nature13461}
  {\bibfield  {journal} {\bibinfo  {journal} {Nature}\ }\textbf {\bibinfo
  {volume} {511}},\ \bibinfo {pages} {202} (\bibinfo {year}
  {2014})}\BibitemShut {NoStop}%
\bibitem [{\citenamefont {Zhang}\ \emph {et~al.}(2017)\citenamefont {Zhang}
  \emph {et~al.}}]{ZhangDPT}%
  \BibitemOpen
  \bibfield  {author} {\bibinfo {author} {\bibfnamefont {J.}~\bibnamefont
  {Zhang}} \emph {et~al.},\ }\href {http://dx.doi.org/10.1038/nature24654}
  {\bibfield  {journal} {\bibinfo  {journal} {Nature}\ }\textbf {\bibinfo
  {volume} {551}},\ \bibinfo {pages} {601} (\bibinfo {year}
  {2017})}\BibitemShut {NoStop}%
\bibitem [{\citenamefont {Ni}\ \emph {et~al.}(2008)\citenamefont {Ni} \emph
  {et~al.}}]{PolarMoleculesExp1}%
  \BibitemOpen
  \bibfield  {author} {\bibinfo {author} {\bibfnamefont {K.-K.}\ \bibnamefont
  {Ni}} \emph {et~al.},\ }\href {\doibase 10.1126/science.1163861} {\bibfield
  {journal} {\bibinfo  {journal} {Science}\ }\textbf {\bibinfo {volume}
  {322}},\ \bibinfo {pages} {231} (\bibinfo {year} {2008})}\BibitemShut
  {NoStop}%
\bibitem [{\citenamefont {Moses}\ \emph {et~al.}(2016)\citenamefont {Moses},
  \citenamefont {Covey}, \citenamefont {Miecnikowski}, \citenamefont {Jin},\
  and\ \citenamefont {Ye}}]{PolarMoleculesExp2}%
  \BibitemOpen
  \bibfield  {author} {\bibinfo {author} {\bibfnamefont {S.~A.}\ \bibnamefont
  {Moses}}, \bibinfo {author} {\bibfnamefont {J.~P.}\ \bibnamefont {Covey}},
  \bibinfo {author} {\bibfnamefont {M.~T.}\ \bibnamefont {Miecnikowski}},
  \bibinfo {author} {\bibfnamefont {D.~S.}\ \bibnamefont {Jin}}, \ and\
  \bibinfo {author} {\bibfnamefont {J.}~\bibnamefont {Ye}},\ }\href@noop {}
  {\bibfield  {journal} {\bibinfo  {journal} {Nature Physics}\ }\textbf
  {\bibinfo {volume} {13}},\ \bibinfo {pages} {13} (\bibinfo {year}
  {2016})}\BibitemShut {NoStop}%
\bibitem [{\citenamefont {Balasubramanian}\ \emph {et~al.}(2009)\citenamefont
  {Balasubramanian} \emph {et~al.}}]{MagneticAtoms}%
  \BibitemOpen
  \bibfield  {author} {\bibinfo {author} {\bibfnamefont {G.}~\bibnamefont
  {Balasubramanian}} \emph {et~al.},\ }\href
  {http://dx.doi.org/10.1038/nature13450} {\bibfield  {journal} {\bibinfo
  {journal} {Nature Materials}\ }\textbf {\bibinfo {volume} {8}},\ \bibinfo
  {pages} {383} (\bibinfo {year} {2009})}\BibitemShut {NoStop}%
\bibitem [{\citenamefont {de~Paz}\ \emph {et~al.}(2013)\citenamefont {de~Paz}
  \emph {et~al.}}]{DipolarGasExp}%
  \BibitemOpen
  \bibfield  {author} {\bibinfo {author} {\bibfnamefont {A.}~\bibnamefont
  {de~Paz}} \emph {et~al.},\ }\href {\doibase 10.1103/PhysRevLett.111.185305}
  {\bibfield  {journal} {\bibinfo  {journal} {Phys. Rev. Lett.}\ }\textbf
  {\bibinfo {volume} {111}},\ \bibinfo {pages} {185305} (\bibinfo {year}
  {2013})}\BibitemShut {NoStop}%
\bibitem [{\citenamefont {Kac}\ and\ \citenamefont
  {Thompson}(1969)}]{KacNormalization}%
  \BibitemOpen
  \bibfield  {author} {\bibinfo {author} {\bibfnamefont {M.}~\bibnamefont
  {Kac}}\ and\ \bibinfo {author} {\bibfnamefont {C.~J.}\ \bibnamefont
  {Thompson}},\ }\href@noop {} {\bibfield  {journal} {\bibinfo  {journal} {J.
  Math. Phys.}\ }\textbf {\bibinfo {volume} {10}},\ \bibinfo {pages} {1373}
  (\bibinfo {year} {1969})}\BibitemShut {NoStop}%
\bibitem [{\citenamefont {Lieb}\ \emph {et~al.}(1961)\citenamefont {Lieb},
  \citenamefont {Schultz},\ and\ \citenamefont {Mattis}}]{LSM}%
  \BibitemOpen
  \bibfield  {author} {\bibinfo {author} {\bibfnamefont {E.}~\bibnamefont
  {Lieb}}, \bibinfo {author} {\bibfnamefont {T.}~\bibnamefont {Schultz}}, \
  and\ \bibinfo {author} {\bibfnamefont {D.}~\bibnamefont {Mattis}},\
  }\href@noop {} {\bibfield  {journal} {\bibinfo  {journal} {Ann. Phys.}\
  }\textbf {\bibinfo {volume} {16}},\ \bibinfo {pages} {407} (\bibinfo {year}
  {1961})}\BibitemShut {NoStop}%
\bibitem [{\citenamefont {Fey}\ and\ \citenamefont
  {Schmidt}(2016)}]{SchmidtLRexcitations}%
  \BibitemOpen
  \bibfield  {author} {\bibinfo {author} {\bibfnamefont {S.}~\bibnamefont
  {Fey}}\ and\ \bibinfo {author} {\bibfnamefont {K.~P.}\ \bibnamefont
  {Schmidt}},\ }\href {\doibase 10.1103/PhysRevB.94.075156} {\bibfield
  {journal} {\bibinfo  {journal} {Phys. Rev. B}\ }\textbf {\bibinfo {volume}
  {94}},\ \bibinfo {pages} {075156} (\bibinfo {year} {2016})}\BibitemShut
  {NoStop}%
\bibitem [{\citenamefont {Vanderstraeten}\ \emph {et~al.}(2018)\citenamefont
  {Vanderstraeten}, \citenamefont {Van~Damme}, \citenamefont {B\"uchler},\ and\
  \citenamefont {Verstraete}}]{VerstraeteLRexcitations}%
  \BibitemOpen
  \bibfield  {author} {\bibinfo {author} {\bibfnamefont {L.}~\bibnamefont
  {Vanderstraeten}}, \bibinfo {author} {\bibfnamefont {M.}~\bibnamefont
  {Van~Damme}}, \bibinfo {author} {\bibfnamefont {H.~P.}\ \bibnamefont
  {B\"uchler}}, \ and\ \bibinfo {author} {\bibfnamefont {F.}~\bibnamefont
  {Verstraete}},\ }\href {\doibase 10.1103/PhysRevLett.121.090603} {\bibfield
  {journal} {\bibinfo  {journal} {Phys. Rev. Lett.}\ }\textbf {\bibinfo
  {volume} {121}},\ \bibinfo {pages} {090603} (\bibinfo {year}
  {2018})}\BibitemShut {NoStop}%
\bibitem [{\citenamefont {Hauke}\ and\ \citenamefont
  {Tagliacozzo}(2013)}]{HaukeTagliacozzo}%
  \BibitemOpen
  \bibfield  {author} {\bibinfo {author} {\bibfnamefont {P.}~\bibnamefont
  {Hauke}}\ and\ \bibinfo {author} {\bibfnamefont {L.}~\bibnamefont
  {Tagliacozzo}},\ }\href {\doibase 10.1103/PhysRevLett.111.207202} {\bibfield
  {journal} {\bibinfo  {journal} {Phys. Rev. Lett.}\ }\textbf {\bibinfo
  {volume} {111}},\ \bibinfo {pages} {207202} (\bibinfo {year}
  {2013})}\BibitemShut {NoStop}%
\bibitem [{\citenamefont {Foss-Feig}\ \emph {et~al.}(2015)\citenamefont
  {Foss-Feig}, \citenamefont {Gong}, \citenamefont {Clark},\ and\ \citenamefont
  {Gorshkov}}]{Gorshkov2}%
  \BibitemOpen
  \bibfield  {author} {\bibinfo {author} {\bibfnamefont {M.}~\bibnamefont
  {Foss-Feig}}, \bibinfo {author} {\bibfnamefont {Z.-X.}\ \bibnamefont {Gong}},
  \bibinfo {author} {\bibfnamefont {C.~W.}\ \bibnamefont {Clark}}, \ and\
  \bibinfo {author} {\bibfnamefont {A.~V.}\ \bibnamefont {Gorshkov}},\ }\href
  {\doibase 10.1103/PhysRevLett.114.157201} {\bibfield  {journal} {\bibinfo
  {journal} {Phys. Rev. Lett.}\ }\textbf {\bibinfo {volume} {114}},\ \bibinfo
  {pages} {157201} (\bibinfo {year} {2015})}\BibitemShut {NoStop}%
\bibitem [{\citenamefont {Buyskikh}\ \emph {et~al.}(2016)\citenamefont
  {Buyskikh}, \citenamefont {Fagotti}, \citenamefont {Schachenmayer},
  \citenamefont {Essler},\ and\ \citenamefont {Daley}}]{DaleyEssler}%
  \BibitemOpen
  \bibfield  {author} {\bibinfo {author} {\bibfnamefont {A.~S.}\ \bibnamefont
  {Buyskikh}}, \bibinfo {author} {\bibfnamefont {M.}~\bibnamefont {Fagotti}},
  \bibinfo {author} {\bibfnamefont {J.}~\bibnamefont {Schachenmayer}}, \bibinfo
  {author} {\bibfnamefont {F.}~\bibnamefont {Essler}}, \ and\ \bibinfo {author}
  {\bibfnamefont {A.~J.}\ \bibnamefont {Daley}},\ }\href {\doibase
  10.1103/PhysRevA.93.053620} {\bibfield  {journal} {\bibinfo  {journal} {Phys.
  Rev. A}\ }\textbf {\bibinfo {volume} {93}},\ \bibinfo {pages} {053620}
  (\bibinfo {year} {2016})}\BibitemShut {NoStop}%
\bibitem [{\citenamefont {\ifmmode \check{Z}\else
  \v{Z}\fi{}unkovi\ifmmode~\check{c}\else \v{c}\fi{}}\ \emph
  {et~al.}(2018)\citenamefont {\ifmmode \check{Z}\else
  \v{Z}\fi{}unkovi\ifmmode~\check{c}\else \v{c}\fi{}}, \citenamefont {Heyl},
  \citenamefont {Knap},\ and\ \citenamefont {Silva}}]{Zunkovic2016}%
  \BibitemOpen
  \bibfield  {author} {\bibinfo {author} {\bibfnamefont {B.}~\bibnamefont
  {\ifmmode \check{Z}\else \v{Z}\fi{}unkovi\ifmmode~\check{c}\else
  \v{c}\fi{}}}, \bibinfo {author} {\bibfnamefont {M.}~\bibnamefont {Heyl}},
  \bibinfo {author} {\bibfnamefont {M.}~\bibnamefont {Knap}}, \ and\ \bibinfo
  {author} {\bibfnamefont {A.}~\bibnamefont {Silva}},\ }\href {\doibase
  10.1103/PhysRevLett.120.130601} {\bibfield  {journal} {\bibinfo  {journal}
  {Phys. Rev. Lett.}\ }\textbf {\bibinfo {volume} {120}},\ \bibinfo {pages}
  {130601} (\bibinfo {year} {2018})}\BibitemShut {NoStop}%
\bibitem [{\citenamefont {Halimeh}\ and\ \citenamefont
  {Zauner-Stauber}(2017)}]{HalimehDPT}%
  \BibitemOpen
  \bibfield  {author} {\bibinfo {author} {\bibfnamefont {J.~C.}\ \bibnamefont
  {Halimeh}}\ and\ \bibinfo {author} {\bibfnamefont {V.}~\bibnamefont
  {Zauner-Stauber}},\ }\href {\doibase 10.1103/PhysRevB.96.134427} {\bibfield
  {journal} {\bibinfo  {journal} {Phys. Rev. B}\ }\textbf {\bibinfo {volume}
  {96}},\ \bibinfo {pages} {134427} (\bibinfo {year} {2017})}\BibitemShut
  {NoStop}%
\bibitem [{\citenamefont {Lerose}\ \emph {et~al.}(2018)\citenamefont {Lerose},
  \citenamefont {Marino}, \citenamefont {Gambassi},\ and\ \citenamefont
  {Silva}}]{LeroseKapitza}%
  \BibitemOpen
  \bibfield  {author} {\bibinfo {author} {\bibfnamefont {A.}~\bibnamefont
  {Lerose}}, \bibinfo {author} {\bibfnamefont {J.}~\bibnamefont {Marino}},
  \bibinfo {author} {\bibfnamefont {A.}~\bibnamefont {Gambassi}}, \ and\
  \bibinfo {author} {\bibfnamefont {A.}~\bibnamefont {Silva}},\ }\href@noop {}
  {\bibfield  {journal} {\bibinfo  {journal} {arXiv:1803.04490}\ } (\bibinfo
  {year} {2018})}\BibitemShut {NoStop}%
\bibitem [{\citenamefont {Else}\ \emph {et~al.}(2017)\citenamefont {Else},
  \citenamefont {Bauer},\ and\ \citenamefont {Nayak}}]{NayakPrethermalTC}%
  \BibitemOpen
  \bibfield  {author} {\bibinfo {author} {\bibfnamefont {D.~V.}\ \bibnamefont
  {Else}}, \bibinfo {author} {\bibfnamefont {B.}~\bibnamefont {Bauer}}, \ and\
  \bibinfo {author} {\bibfnamefont {C.}~\bibnamefont {Nayak}},\ }\href
  {\doibase 10.1103/PhysRevX.7.011026} {\bibfield  {journal} {\bibinfo
  {journal} {Phys. Rev. X}\ }\textbf {\bibinfo {volume} {7}},\ \bibinfo {pages}
  {011026} (\bibinfo {year} {2017})}\BibitemShut {NoStop}%
\bibitem [{\citenamefont {Jurcevic}\ \emph {et~al.}(2017)\citenamefont
  {Jurcevic} \emph {et~al.}}]{JurcevicDQPT}%
  \BibitemOpen
  \bibfield  {author} {\bibinfo {author} {\bibfnamefont {P.}~\bibnamefont
  {Jurcevic}} \emph {et~al.},\ }\href {\doibase 10.1103/PhysRevLett.119.080501}
  {\bibfield  {journal} {\bibinfo  {journal} {Phys. Rev. Lett.}\ }\textbf
  {\bibinfo {volume} {119}},\ \bibinfo {pages} {080501} (\bibinfo {year}
  {2017})}\BibitemShut {NoStop}%
\bibitem [{\citenamefont {Haegeman}\ \emph {et~al.}(2016)\citenamefont
  {Haegeman}, \citenamefont {Lubich}, \citenamefont {Oseledets}, \citenamefont
  {Vandereycken},\ and\ \citenamefont {Verstraete}}]{haegeman2016unifying}%
  \BibitemOpen
  \bibfield  {author} {\bibinfo {author} {\bibfnamefont {J.}~\bibnamefont
  {Haegeman}}, \bibinfo {author} {\bibfnamefont {C.}~\bibnamefont {Lubich}},
  \bibinfo {author} {\bibfnamefont {I.}~\bibnamefont {Oseledets}}, \bibinfo
  {author} {\bibfnamefont {B.}~\bibnamefont {Vandereycken}}, \ and\ \bibinfo
  {author} {\bibfnamefont {F.}~\bibnamefont {Verstraete}},\ }\href {\doibase
  10.1103/PhysRevB.94.165116} {\bibfield  {journal} {\bibinfo  {journal} {Phys.
  Rev. B}\ }\textbf {\bibinfo {volume} {94}},\ \bibinfo {pages} {165116}
  (\bibinfo {year} {2016})}\BibitemShut {NoStop}%
\bibitem [{\citenamefont {Haegeman}\ \emph {et~al.}(2011)\citenamefont
  {Haegeman}, \citenamefont {Cirac}, \citenamefont {Osborne}, \citenamefont
  {Pi\ifmmode~\check{z}\else \v{z}\fi{}orn}, \citenamefont {Verschelde},\ and\
  \citenamefont {Verstraete}}]{haegeman2011time}%
  \BibitemOpen
  \bibfield  {author} {\bibinfo {author} {\bibfnamefont {J.}~\bibnamefont
  {Haegeman}}, \bibinfo {author} {\bibfnamefont {J.~I.}\ \bibnamefont {Cirac}},
  \bibinfo {author} {\bibfnamefont {T.~J.}\ \bibnamefont {Osborne}}, \bibinfo
  {author} {\bibfnamefont {I.}~\bibnamefont {Pi\ifmmode~\check{z}\else
  \v{z}\fi{}orn}}, \bibinfo {author} {\bibfnamefont {H.}~\bibnamefont
  {Verschelde}}, \ and\ \bibinfo {author} {\bibfnamefont {F.}~\bibnamefont
  {Verstraete}},\ }\href {\doibase 10.1103/PhysRevLett.107.070601} {\bibfield
  {journal} {\bibinfo  {journal} {Phys. Rev. Lett.}\ }\textbf {\bibinfo
  {volume} {107}},\ \bibinfo {pages} {070601} (\bibinfo {year}
  {2011})}\BibitemShut {NoStop}%
\bibitem [{Note1()}]{Note1}%
  \BibitemOpen
  \bibinfo {note} {At finite $L$, due to inversion symmetry with respect to the
  center of the open chain, the eigenstates of $H_{\protect \text {eff}}$ are
  actually even and odd superpositions of wavefunctions localized at symmetric
  positions in the chain. However, such a finite-size effect is unstable to any
  perturbation which breaks this symmetry, such as, e.g., random boundary
  conditions.}\BibitemShut {Stop}%
\bibitem [{\citenamefont {Halimeh}\ \emph {et~al.}(2018)\citenamefont
  {Halimeh}, \citenamefont {Van~Damme}, \citenamefont {Zauner-Stauber},\ and\
  \citenamefont {Vanderstraeten}}]{Halimeh:ADQPTandConfinement}%
  \BibitemOpen
  \bibfield  {author} {\bibinfo {author} {\bibfnamefont {J.~C.}\ \bibnamefont
  {Halimeh}}, \bibinfo {author} {\bibfnamefont {M.}~\bibnamefont {Van~Damme}},
  \bibinfo {author} {\bibfnamefont {V.}~\bibnamefont {Zauner-Stauber}}, \ and\
  \bibinfo {author} {\bibfnamefont {L.}~\bibnamefont {Vanderstraeten}},\
  }\href@noop {} {\bibfield  {journal} {\bibinfo  {journal} {arxiv:1810.07187}\
  } (\bibinfo {year} {2018})}\BibitemShut {NoStop}%
\bibitem [{\citenamefont {Mori}(2018)}]{mori2018prethermalization}%
  \BibitemOpen
  \bibfield  {author} {\bibinfo {author} {\bibfnamefont {T.}~\bibnamefont
  {Mori}},\ }\href@noop {} {\bibfield  {journal} {\bibinfo  {journal} {arXiv
  preprint arXiv:1810.01584}\ } (\bibinfo {year} {2018})}\BibitemShut {NoStop}%
\bibitem [{\citenamefont {Rutkevich}(2008)}]{RutkevichSemiclassical}%
  \BibitemOpen
  \bibfield  {author} {\bibinfo {author} {\bibfnamefont {S.}~\bibnamefont
  {Rutkevich}},\ }\href {https://doi.org/10.1007/s10955-008-9495-1} {\bibfield
  {journal} {\bibinfo  {journal} {J. Stat. Phys.}\ }\textbf {\bibinfo {volume}
  {131}},\ \bibinfo {pages} {917} (\bibinfo {year} {2008})}\BibitemShut
  {NoStop}%
\bibitem [{\citenamefont {Neyenhuis}\ \emph {et~al.}(2017)\citenamefont
  {Neyenhuis} \emph {et~al.}}]{NeyenhuisPrethermalizationLRExp}%
  \BibitemOpen
  \bibfield  {author} {\bibinfo {author} {\bibfnamefont {B.}~\bibnamefont
  {Neyenhuis}} \emph {et~al.},\ }\href {\doibase 10.1126/sciadv.1700672}
  {\bibfield  {journal} {\bibinfo  {journal} {Science Advances}\ }\textbf
  {\bibinfo {volume} {3}} (\bibinfo {year} {2017}),\
  10.1126/sciadv.1700672}\BibitemShut {NoStop}%
\bibitem [{\citenamefont {Schachenmayer}\ \emph {et~al.}(2013)\citenamefont
  {Schachenmayer}, \citenamefont {Lanyon}, \citenamefont {Roos},\ and\
  \citenamefont {Daley}}]{Daley}%
  \BibitemOpen
  \bibfield  {author} {\bibinfo {author} {\bibfnamefont {J.}~\bibnamefont
  {Schachenmayer}}, \bibinfo {author} {\bibfnamefont {B.~P.}\ \bibnamefont
  {Lanyon}}, \bibinfo {author} {\bibfnamefont {C.~F.}\ \bibnamefont {Roos}}, \
  and\ \bibinfo {author} {\bibfnamefont {A.~J.}\ \bibnamefont {Daley}},\ }\href
  {\doibase 10.1103/PhysRevX.3.031015} {\bibfield  {journal} {\bibinfo
  {journal} {Phys. Rev. X}\ }\textbf {\bibinfo {volume} {3}},\ \bibinfo {pages}
  {031015} (\bibinfo {year} {2013})}\BibitemShut {NoStop}%
\bibitem [{\citenamefont {Lerose}\ and\ \citenamefont
  {Pappalardi}(2018)}]{LeroseEE}%
  \BibitemOpen
  \bibfield  {author} {\bibinfo {author} {\bibfnamefont {A.}~\bibnamefont
  {Lerose}}\ and\ \bibinfo {author} {\bibfnamefont {S.}~\bibnamefont
  {Pappalardi}},\ }\href@noop {} {\bibfield  {journal} {\bibinfo  {journal}
  {arXiv:1811.05505}\ } (\bibinfo {year} {2018})}\BibitemShut {NoStop}%
\bibitem [{\citenamefont {James}\ \emph {et~al.}(2018)\citenamefont {James},
  \citenamefont {Konik},\ and\ \citenamefont {Robinson}}]{KonikConfinement}%
  \BibitemOpen
  \bibfield  {author} {\bibinfo {author} {\bibfnamefont {A.~J.~A.}\
  \bibnamefont {James}}, \bibinfo {author} {\bibfnamefont {R.~M.}\ \bibnamefont
  {Konik}}, \ and\ \bibinfo {author} {\bibfnamefont {N.~J.}\ \bibnamefont
  {Robinson}},\ }\href@noop {} {\bibfield  {journal} {\bibinfo  {journal}
  {arxiv:1804.09990}\ } (\bibinfo {year} {2018})}\BibitemShut {NoStop}%
\bibitem [{\citenamefont {Lin}\ and\ \citenamefont
  {Motrunich}(2017)}]{LinMotrunich:Quasiconserved}%
  \BibitemOpen
  \bibfield  {author} {\bibinfo {author} {\bibfnamefont {C.-J.}\ \bibnamefont
  {Lin}}\ and\ \bibinfo {author} {\bibfnamefont {O.~I.}\ \bibnamefont
  {Motrunich}},\ }\href {\doibase 10.1103/PhysRevB.96.214301} {\bibfield
  {journal} {\bibinfo  {journal} {Phys. Rev. B}\ }\textbf {\bibinfo {volume}
  {96}},\ \bibinfo {pages} {214301} (\bibinfo {year} {2017})}\BibitemShut
  {NoStop}%
\end{thebibliography}%
\end{document}